# Research on Brick Schema Representation for Building Operation with Variable Refrigerant Flow Systems

Jingming Li[a, b], Nianping Li[a, *], Rui Yan[b], Kushnazarov Farruh[b], Anbang Li[b], Kehua Li[b]

[a]College of Civil Engineering, Hunan University, Changsha 410082, China

[a]HVAC & Building Technologies Division, Midea Group, Foshan, 528311, China

**Abstract**

Building metadata is regarded as the signpost in organizing massive building data. The application of building metadata simplifies the creation of digital representations and provides portable data analytics. Typical metadata standards such as Brick and Haystack are used to describe the data of the building system. Brick uses standard ontologies to create building metadata. However, neither Haystack nor Brick has provided definitions about the Variable Refrigerant Flow (VRF) system so far. For years, both Brick and Haystack working groups have been discussing how to describe VRF in their schema, mainly about the classification of VRF and the definitions of VRF units. There were no settled solutions for these problems. Meanwhile, the global VRF market is growing increasingly fast because of the energy efficiency and installation simplicity of the VRF system. It is needed to have the metadata to describe VRF units in buildings for data analysis and management. Addressing this challenge, this paper extended Brick Schema with the VRF module and verified the Brick VRF module. Then, the model and the service framework were developed and applied for a building in China. The framework can serve portable energy analysis for different areas. The VRF module of this paper provides a possible solution for the expression of the VRF system in the building semantic web. The works in this paper will support semantic web in automation strategies for building management and scalable building operation.

**Keywords:** Brick; VRF; Building Metadata; Building management; Ontology

# 1. Introduction

Buildings have interdependent systems and components for various purposes, such as building operation, indoor environment quality, security, and Heating, Ventilation, and Air Conditioning (HVAC). More than the one-third energy consumption of commercial buildings is from HVAC (U.S. Energy Information Administration 2020). Among them, more than 80% of commercial buildings in Europe and East Asia use Variable Refrigerant Flow (VRF) systems for HVAC (Joanna R. Turpin 2017). Compared with the traditional HVAC devices, the VRF system has advantages in satisfying part-load performance, flexible individual control, no duct delivery loss, and easy installation and maintenance (Zhang et al. 2019; Zhao, Zhang, and Zhong 2015). The simplification of the pipelines benefits from the ingenious design of the indoor and outdoor units. Consequently, the indoor and outdoor units of the VRF system are very sophisticated, because nearly all the sensors, valves, and control units are integrated inside them. Meanwhile, the VRF is undergoing an accelerated growth. The global VRF market is expected to reach USD 30.89 billion by 2025 (Verified Market Research 2018). The complicity and popularity of the VRF system make it crucial to the successful management of building energy.

To precisely and properly manage building energy, smart building, and digital twin have gained attention (Metallidou, Psannis, and Egyptiadou 2020; Teng et al. 2021). Smart building requires effective coordination and even integration of these independent systems, which has become the mainstream for collaborative operation and maintenance of assets and infrastructures. The smart building relies on correct and precise digital reflection of the real world using digital twin, making the access and analysis of data a great challenge.

Haystack was introduced to deal with this challenge. It provides building components definition and relationship deductions through tags. It brought the concept of Tag to building metadata, which is an effective method to represent, transfer, and explain data. A tag is a name/value pair applied to an entity. For instance, the id tag representing an entity has a Ref value in (id: @whitehouse). Tags represent the facts about data items and can be associated with roll calls to provide information describing objects. Also, Haystack brought standard and extendable libraries of tags. It has model implantations, application specifications, and data serialization formats. The emergence of Haystack provides a way for building systems and intelligent devices to implement automatic interpretation and common vocabulary in various software and web-based applications. Although Haystack intended to streamline working with data from the Internet of Things through standardization and externality tagging labels, there are problems during its application. Bergmann et al. (2020) summarized several issues of Haystack. They pointed out that Haystack lacks verification, due to its vast definition of models, applications, and formats. The models in Haystack lack readability, they are neither articulated nor machine-interpretable. Another issue is that Haystack did not provide example applications or manuals for applying the schema.

Brick inherited the ideas from Haystack and described metadata with Entity, Class, Relationship, and Graph (Balaji et al. 2018). Brick intends to connect all sorts of building data through the building's lifecycle (Gabe Fierro, Prakash, et al. 2020). Comparing with Haystack, Brick demonstrated better visualization and readability (Gabe Fierro, Koh, et al. 2020). Balaji et al. (2018) demonstrated Brick can map up to nearly all Building Management System (BMS) data points across six buildings. Google has applied its digital building platform to more than 100 buildings with Brick as the resource description framework. Brick (Gabe Fierro et al. 2019) also provided an open test-bed for building operation data analysis.

Brick is trending, regarding building metadata representation. However, Brick has not supported modeling the VRF system since being promoted. Regarding cooling/heating system, brick schema only covered water-based system. The Brick group has been discussing the development of VRF for a while (Clement Bouvier Europe et al. 2020). The first thing to consider is that there are no definitions of VRF or Hydrofluorocarbon in the schema. The dilemma is how to classify the VRF. The discussion about whether its classification is an Air Handling Unit, or a Chiller, or a Heat Pump has always been inconclusive. Because the roles of the VRF components change as the working mode changes.

Therefore, this study develops the VRF system in Brick Schema. The Brick VRF module is validated and applied to an office building. The paper is organized as follows, Section 2 explains the background of adopting Brick for managing VRF along with a comparison of the latest version of Brick Schema and Project Haystack. Section 3 presented the development of the VRF module in Brick. The application of VRF Brick modules in the BMS is demonstrated in Section 4. These are followed by a discussion in Section 5. Section 6 concludes the paper.

**2. Brick for BMS**

As mentioned above, Haystack has been criticized for a long time because of the lack of consistency of the system, which is the reason that Brick is trending. A comparison on model representation indicated that the performance of Brick (Version 1.1.0) was higher than Haystack (Version 3.9.7), regarding the model completeness and the relationship expressiveness (Quinn and McArthur 2021). However, since the cooperation of ASHRAE's BACnet Committee (Haynes 2018), the two systems have been continuously updating versions, absorbing the advantages of each other. Therefore, this paper compared the latest versions of Haystack (Version 3.9.10, referred to as Haystack 4) and Brick (Version 1.2.0).

**2.1. A Comparison of Brick and Haystack**

The design of Brick took the completeness of all information, the expressivity in relationships, the usability for various users, and consistency along with the model processes, and the extensibility for more concepts.

The expression of relations at all levels is consistent in Brick (Brick Schema 2021). Figure 1 showed example is a chiller in Brick:

```
brick:Chiller a owl:Class ;
    rdfs:label "Chiller" ;
    rdfs:subClassOf [ owl:intersectionOf ( _:has_Equipment _:has_Chiller ) ],
        brick:HVAC_Equipment ;
    skos:definition "Refrigerating machine used to transfer heat between fluids. Chillers are
either direct expansion with a compressor or absorption type."@en ;
    brick:hasAssociatedTag tag:Chiller,
        tag:Equipment .
```

**Fig. 1.** An example of the model in Brick Schema

As mentioned above, the codes showed their parent and sub connections. The chiller is built with RDF and OWL from W3C. Brick is more user-friendly and provides a guide from getting started to mastering in terms of application and development, as well as the best-case applications. For the moment, Brick has not provided integrations with Industry Foundation Classes (IFC) for Building Information Modeling (BIM) applications. But there is Brick based digital building platform working on integrating IFC.

```
phIoT:chiller a owl:Class ;
  rdfs:subClassOf phIoT:equip ;
  rdfs:label "chiller" ;
  ph:children "ZincDict(\"{chilled water delta flow sensor point}\")"^^ph:xstr,
    "ZincDict(\"{chilled water delta temp sensor point}\")"^^ph:xstr,
    "ZincDict(\"{chilled water entering pipe equip}\")"^^ph:xstr,
    "ZincDict(\"{chilled water leaving pipe equip}\")"^^ph:xstr,
    "ZincDict(\"{chilled water valve isolation cmd point}\")"^^ph:xstr,
    "ZincDict(\"{condenser refrig pressure sensor point}\")"^^ph:xstr,
    "ZincDict(\"{condenser refrig temp sensor point}\")"^^ph:xstr,
    "ZincDict(\"{condenser run state point}\")"^^ph:xstr,
    "ZincDict(\"{condenser water entering pipe equip}\")"^^ph:xstr,
    "ZincDict(\"{condenser water leaving pipe equip}\")"^^ph:xstr,
    "ZincDict(\"{condenser water valve isolation cmd point}\")"^^ph:xstr,
    "ZincDict(\"{efficiency sensor point}\")"^^ph:xstr,
    "ZincDict(\"{enable state point}\")"^^ph:xstr,
    "ZincDict(\"{equip}\")"^^ph:xstr,
    "ZincDict(\"{evaporator refrig pressure sensor point}\")"^^ph:xstr,
    "ZincDict(\"{evaporator refrig temp sensor point}\")"^^ph:xstr,
    "ZincDict(\"{load cmd point}\")"^^ph:xstr,
    "ZincDict(\"{load sensor point}\")"^^ph:xstr,
    "ZincDict(\"{point}\")"^^ph:xstr,
    "ZincDict(\"{run state point}\")"^^ph:xstr ;
  phIoT:chillerMechanism phIoT:chillerMechanismType ;
  phIoT:cools phIoT:chilled-water ;
  rdfs:comment "Equipment to remove heat from a liquid.  Chillers typically use a vapor\ncompression or an absorption refrigeration cycle." ;
  ph:is phIoT:equip ;
  ph:lib phIoT:lib:phIoT ;
    ph:wikipedia "https://en.wikipedia.org/wiki/Chiller"^^xsd:anyURI .
```

**Fig. 2.** An example of the model in Project Haystack

Haystack 4.0 (Project Haystack 2021), added several new features. Figure 2 is an example of a chiller in Haystack. It has adopted Web Ontology Language (OWL) and RDF for ontology. The Namespace, Def are declared as Point 3 in the above summary. It is needed to notice that the chiller in Turtle format referred to the classes defined in Zinc.

The new version shows three real cases of the Haystack application, which is helpful for novices to understand and develop the protocol. Haystack has added new features to compensate for its inconsistency. However, Namespace, Def, Subtyping, Reflection, and Relationship have functional overlaps. For instance, water is a subtype of fluid, which can be described as isPartOf in Brick. It is unnecessary to classify tags as libraries. The other issue is that the new version failed to provide a tutorial for starters, nor a testbed for education. Haystack provides all the tools to the users, regardless of their ability and knowledge in the field. Therefore, considering the model consistency and the ease of use of the application, the semantic development of the VRF module is based on Brick.

## 2.2. The role of Brick Schema in BMS

The relationship defined in the Brick model represents the connection between the datasets. As shown in Figure 3, the role of Brick is to create reliable connections in the data lake, assisting the digital twin process from devices to application interfaces in BACnet.

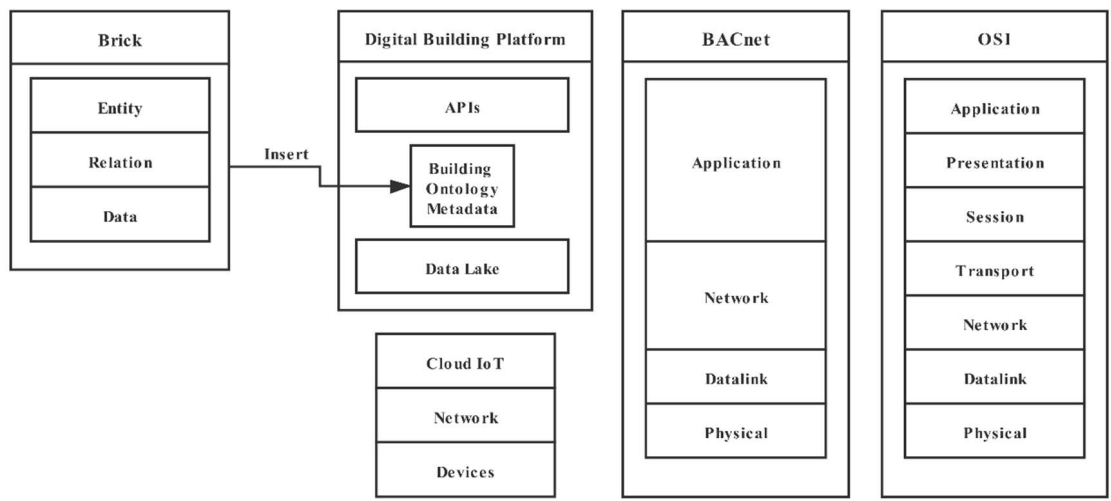

**Fig. 3.** The role of Brick Schema in BMS

To fulfill its role in BMS, the developers designed the framework for instant queries in massive databases as indicated in Figure 4 (Gabe Fierro et al. 2019). The request is firstly verified against the relationships in a Resource Description Framework (RDF) database, which forms all relevant queries for the timeseries database. The timeseries database, then, return matched tables and deliver the results.

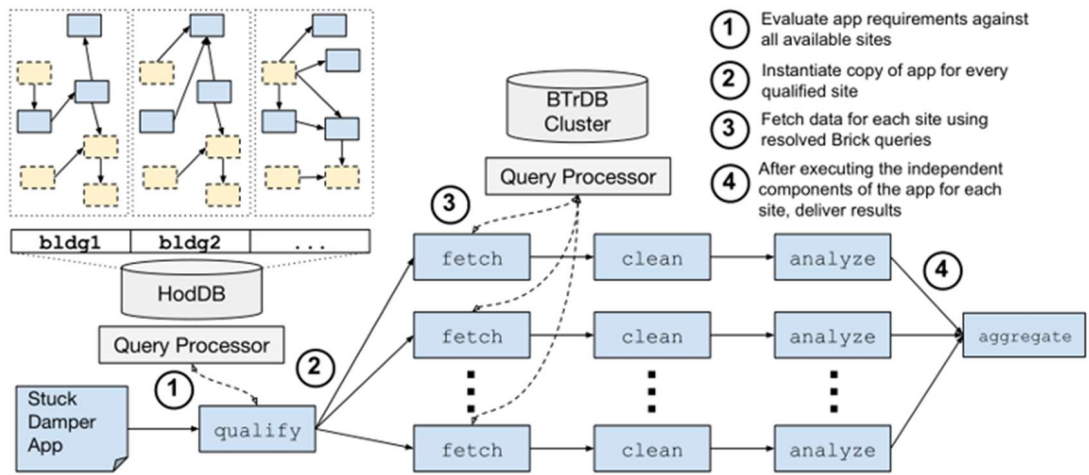

**Fig. 4.** The framework of a server using Brick (Gabe Fierro et al. 2019)

The intention of developing the VRF module in Brick is to link timetables in BMS databases with VRF devices. Relational definitions as indexes for database accesses can provide a basis for more effective executions of multi-database queries during building operations.

**3. Developing Brick Tags for VRF**

Brick and its branches have been used for hundreds of buildings. However, bringing up a VRF module has been a problem in Haystack (Josiah and Frank 2019) and similar discussions happen in Brick Group. The discussion on how to represent VRF in Brick has been unsettled since 2020. For a water-cooling system, the names are defined based on their position during the loop. For instance, in Figure 5, 'Entering Water Temperature Sensor' means a temperature sensor for water entering a condenser. 'Heat Exchanger Supply Water Temperature Sensor' measures the temperature of water supplied by a heat exchanger. Such naming clearly expresses the role and location of the equipment in the water system, but it is not appropriate in the VRF system.

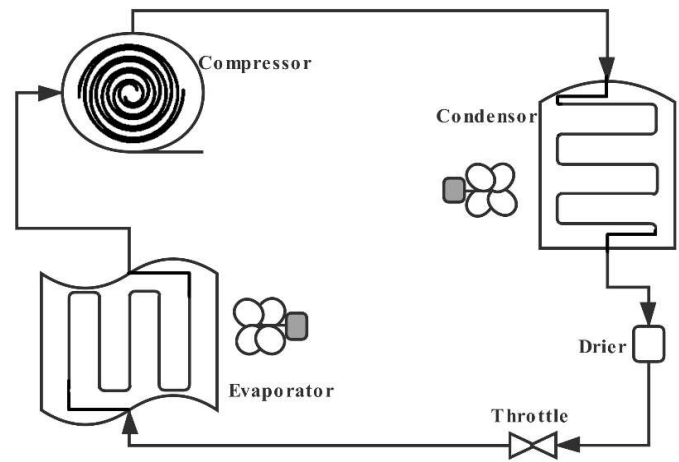

**Fig. 5.** A Water Cooling System

**3.1.** Developing Brick for VRF.

The cooling and heating loops are demonstrated in Figure 6. The four-way valve plays an important role in VRF systems. When cooling, Port A links to Port D and Port B links to Port C, the direction of refrigerant flow is the same as the direction of water flow. When heating, Port A links to Port B and Port C links to Port D, the general direction of refrigerant flow has been reversed. If a VRF device for Brick (such as a temperature sensor) is defined as an 'Entering Refrigerant Temperature Sensor' during the cooling loop. The name can hardly fit its status during the heating loop.

Since the status of the device changes considering their working conditions, the development of the VRF Brick module used the device roles during cooling mode as the device names and the positions to represent the sensors. The data transferred from the devices can distinguish their working condition. The development work covers everything from the definition of refrigerant to the definition of VRF equipment. As for the sensors such as frequency and pressure, Brick has already included sensors for HVAC, which the developed Brick VRF can also use. Figure 7 showed some of the created modules. After compiling, the Brick VRF module can be applied for projects.

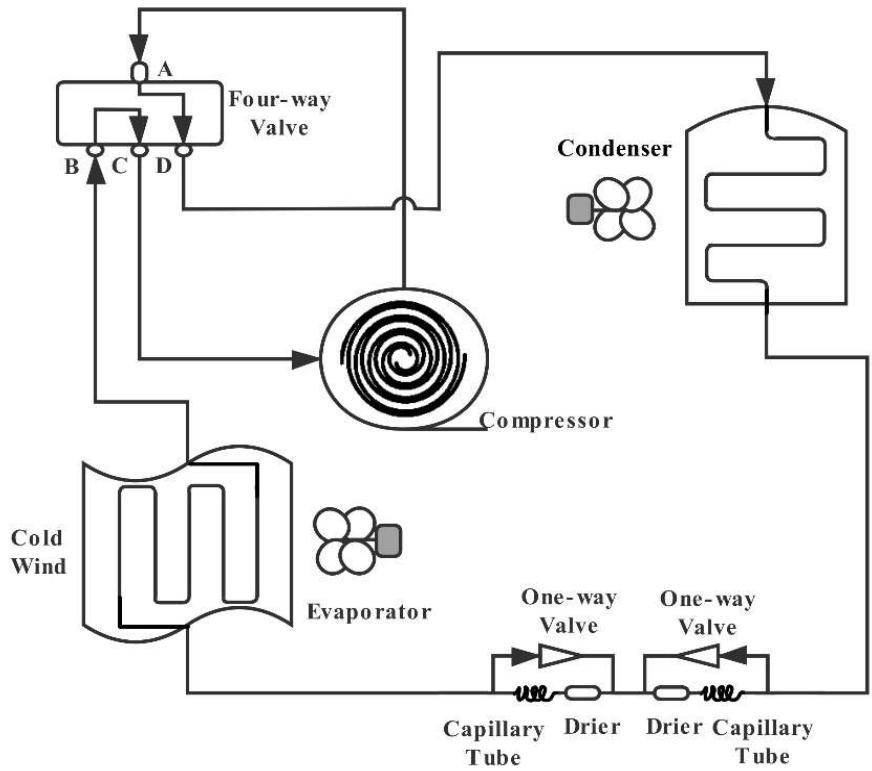

**(a)** VRF cooling loop

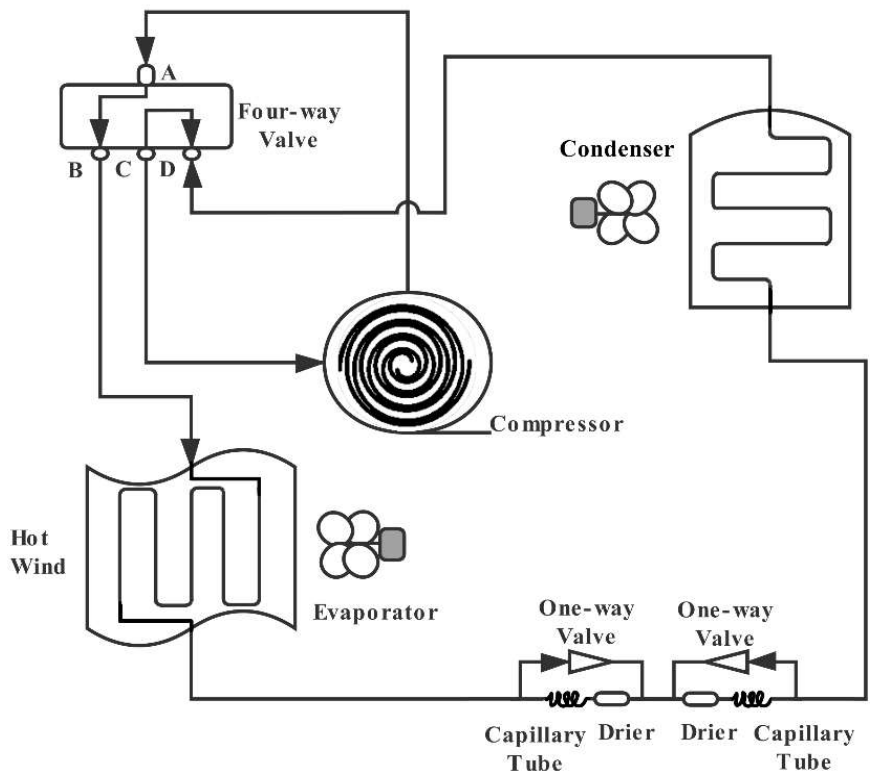

**(b)** VRF heating loop

**Fig. 6.** A VRF System.

```
"Hydrofluorocarbon_System": {            "Four_way_Valve": {
        "tags": [                                "tags": [
          TAG.Hydrofluorocarbon,                   TAG.Gas, TAG.Valve, TAG.Equipment
          TAG.System,                            ]},
        ]
      },
```

  **(a) Define Hydrofluorocarbon System**  **(b) Define Four-way Valve**

```
"Hydrofluorocarbon_Gas": {
          "tags": [TAG.Fluid, TAG.Gas, TAG.Hydrofluorocarbon_Gas],
          "subclasses": {
            "R410A_Gas": {
               "tags": [TAG.Fluid, TAG.Gas, TAG.Hydrofluorocarbon_Gas, TAG.R410A_Gas],
            },
          },
        },
```

  **(c) Define Subtance - Hydrofluorocarbon Gas**

```
"Header": {"tags": [TAG.Gas, TAG.Liquid, TAG.Distribution, TAG.Equipment]}
```

  **(d) Define Header**

```
"Separation_Tube": {"tags": [TAG.Gas, TAG.Liquid, TAG.Distribution, TAG.Equipment]}
```

  **(e) Define Separation Tube**

```
"Hydrofluorocarbon": {
          "tags": [TAG.Fluid, TAG.Liquid, TAG.Hydrofluorocarbon],
          "subclasses": {
            "R410A": {
               "tags": [TAG.Fluid, TAG.Liquid, TAG.Hydrofluorocarbon, TAG.R410A]
            },
          },
        },
```

  **(f) Define Subtance - Hydrofluorocarbon**

```
"VRF_Outdoor": {                          "VRF_Indoor": {
    "tags": [                                 "tags": [
      TAG.Equipment,                            TAG.Equipment,
      TAG.VRF_Outdoor,                          TAG.VRF_Indoor,
      TAG.Heat_Exchanger,                       TAG.Heat_Exchanger,
      TAG.Fan,                                  TAG.Fan,
      TAG.Compressor,                           TAG.Cool,
      TAG.Check_Valve,                          TAG.Heat,
      TAG.Four_way_Valve,                       TAG.Hydrofluorocarbon,
      TAG.Cool,                                 TAG.Electromagnetic_Valve,
      TAG.Heat,                                 TAG.Volume,
      TAG.Hydrofluorocarbon,                    TAG.Box,
      TAG.Supply,                               TAG.Supply,
      TAG.Return,                               TAG.Return,
      TAG.Point,                                TAG.Setpoint,
      TAG.Pressure,                             TAG.Point,
      TAG.Setpoint,                             TAG.sensor,
      TAG.Electronic_Expansion_Valve,           TAG.Electronic_Expansion_Valve,
      TAG.sensor,                             ],
    ],                                      },
  },
}
```

  **(g) Define VRF Outdoor Unit**  **(h) Define VRF Intdoor Unit**

**Fig. 7.** Partial Codes of the developed VRF module

## 3.2. Representing VRF in Brick

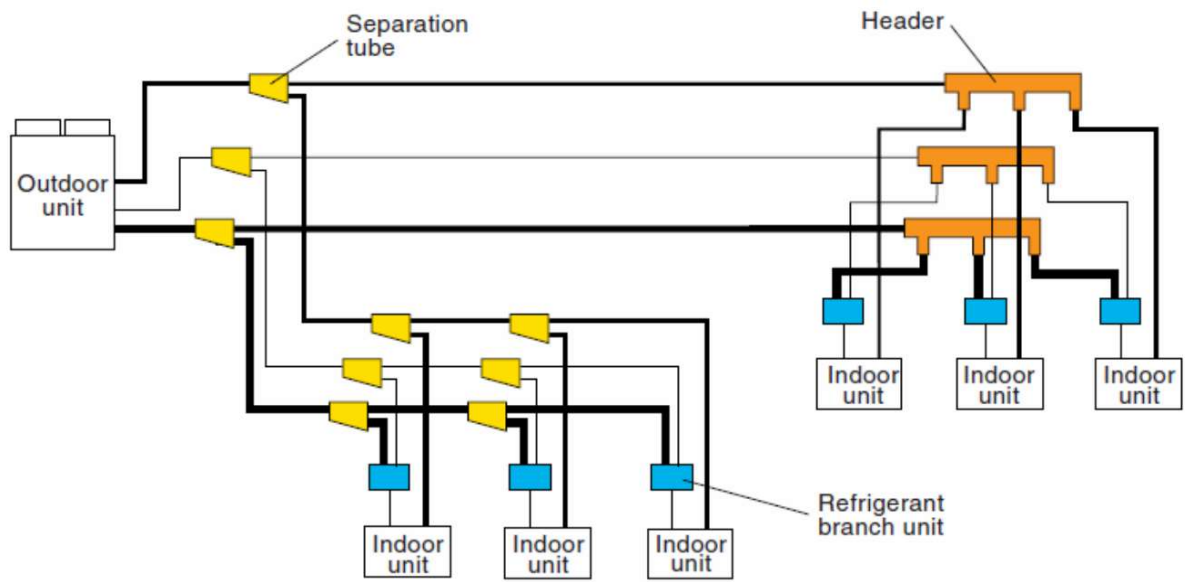

**Fig. 8.** Heat Recovery Type VRF System (Bhatia 2011)

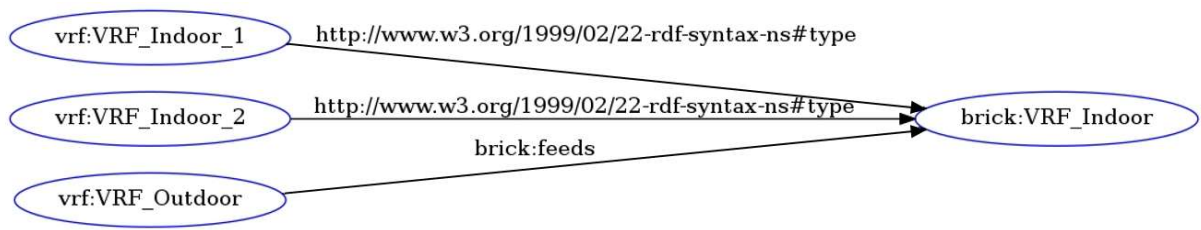

**Fig.9.** Using symmetric relationship to describe VRF

The extended VRF module of Brick can help describe the devices semanticly. For instance, the heat recovery system in Figure 8 uses a three-pipe system for the liquid line, the hot gas line, and the suction line. Each indoor unit is connected to the three pipes through solenoid valves, the indoor unit opens the liquid and the suction pipes to serve as a evaporator when cooling, it opens the liquid and hot gas pipes to serve as a condenser when cooling. Although the circulation direction of the refrigents are different in cooling and heating loops, the indoor units are still fed by outdoor units, the heat revovery VRF system can be expressed by 'feeds' in Figure 9.

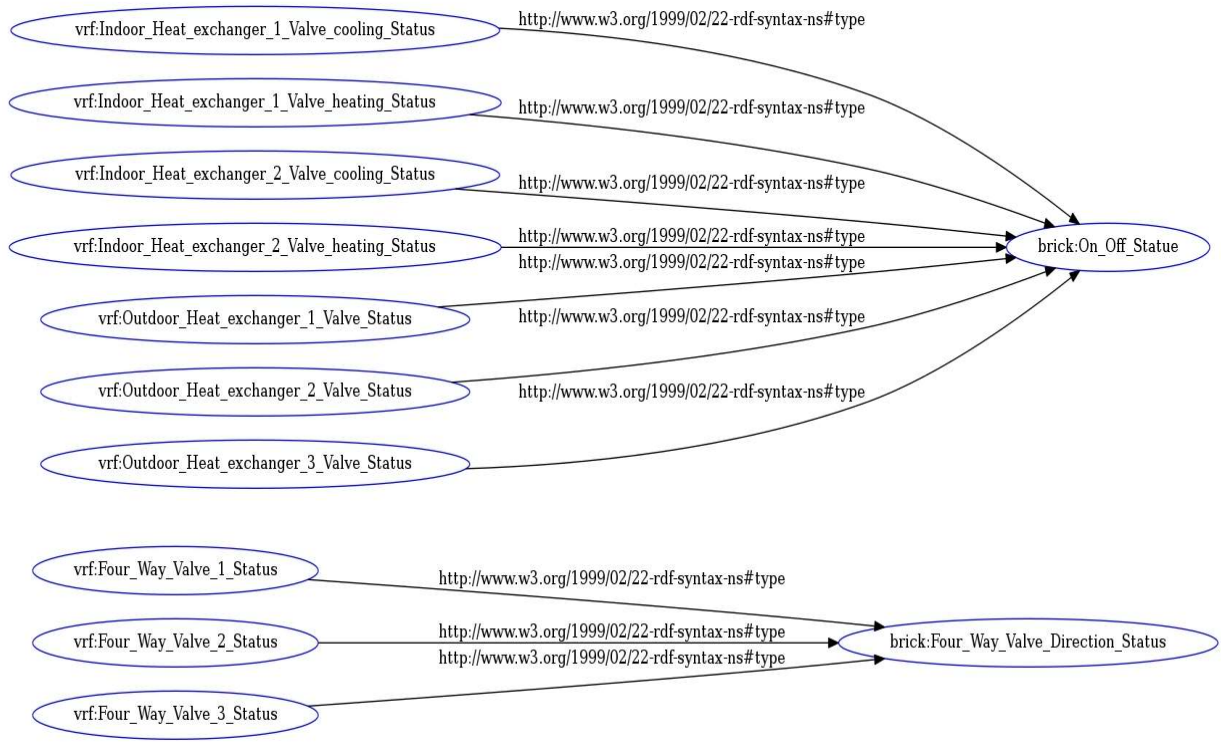

**Fig.10.** Using status of sensors to describe VRF working condition

The internal structures can also be described semanticaly. In Figure 10, the status of valves can determine the working conditions, the on-off status of can determine whether the working mode is cooling, heating or both.

### 3.3. BMS server with Brick VRF module

A server with databases to store the building semantic model and building operation data is deployed. The server is developed upon Brick-server (Gabriel Fierro 2019; Schneider et al. 2020). As illustrated in Figure 11, the sensor data goes to databases through a gateway. The building model with VRF systems is stored in the databases, data labeled with timestamps are stored in timescale databases corresponding to their points in the model. The server in Figure 11 takes advantage of Brick Schema in portable building analysis with various buildings (Gabe Fierro and Culler 2017; Koh et al. 2018). Clients can query the database for building relationships, as well as fetch data for operation analysis.

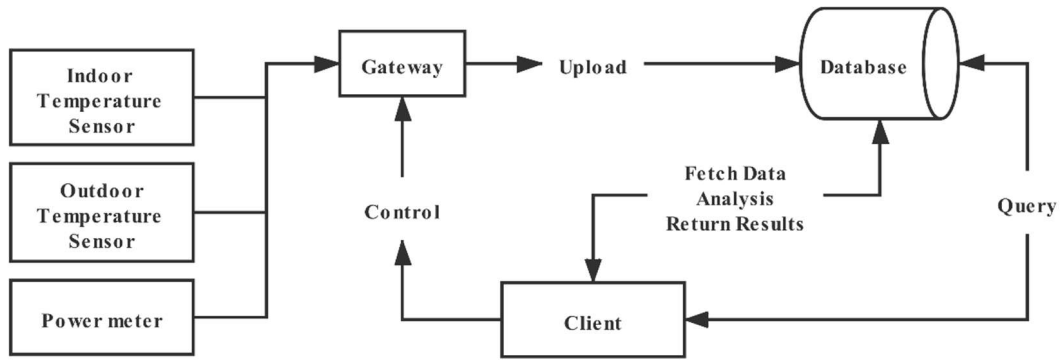

**Fig. 11.** BMS data server with Brick VRF module

## 3.4. Validation

To verify the Brick VRF module, an experimental model is built for validation. Figure 12 illustrates the internal relationships of the apartment building. Apart from sensors, an HVAC zone fed by a VRF system is also included.

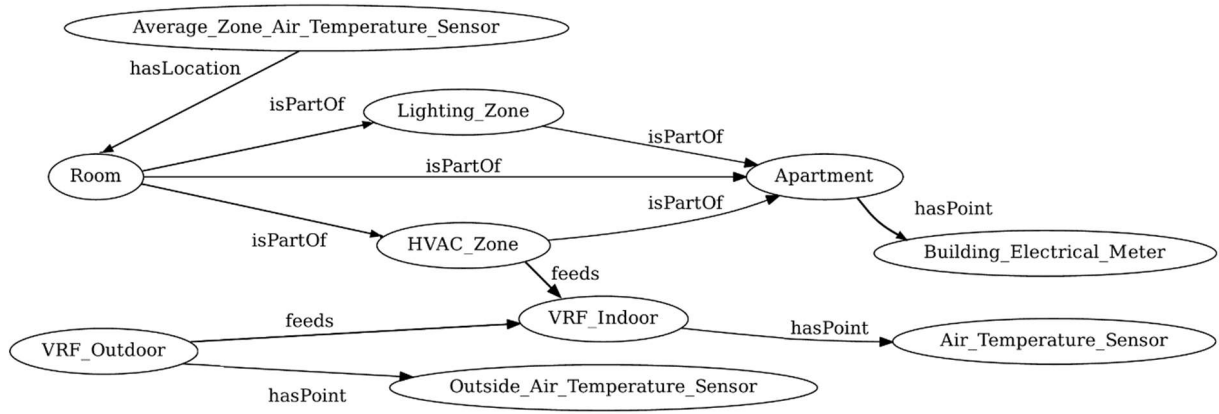

**Fig. 12.** The model for validating the Brick VRF model

After building the model from source codes, it is uploaded to the server to test if the relationships with the developed VRF module are compatible with queries. The validation results regarding a query are in Figure 13. The query returned all results related to Room, including the VRF indoor and outdoor units. The Brick VRF module is considered validated.

Fig. 13. Evaluation with SPARQL query.

The last test is to validate if the BMS data server could host data manipulation with the VRF model. The open data from Mortar (Gabe Fierro et al. 2019) is used to test the performance of the server. The gateway reads and uploads the downloaded data to the server in the format of [id, time, value]. The test sends queries to the server, and the server will filter and return the data of the objects that meet the requirements. The advantage of this server is that the BMS data management is no longer limited to a single building, and the building data that fits the same ontology standard can be queried and analyzed, instantly. Figure 14 showed the test results, and energy balance analysis was done simultaneously when getting energy and outside temperature data from the query.

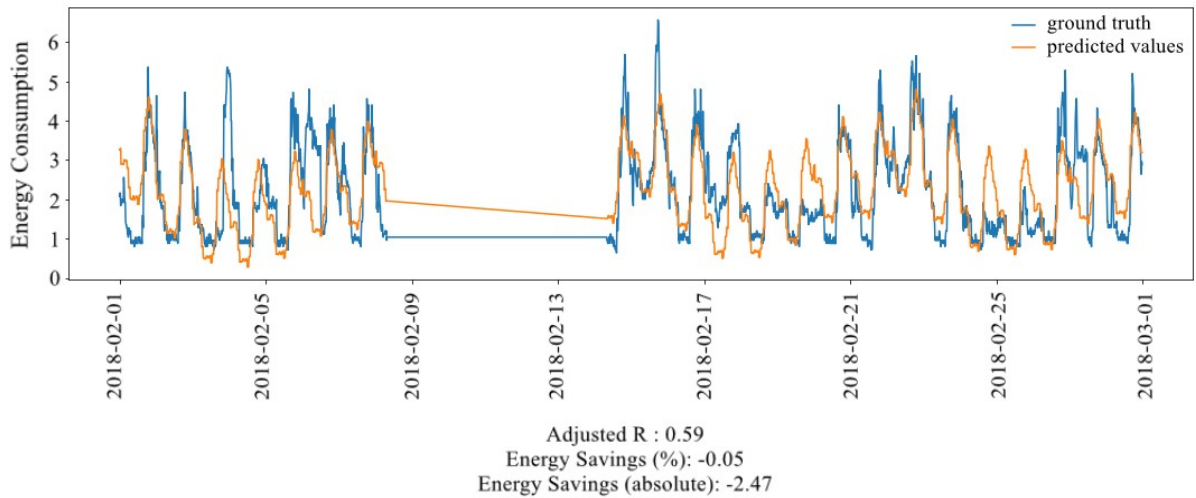

**Fig. 14.** Instant analysis of the returned data

## 4. Application

After the Brick VRF module and the server being validated, they are used for a building sector in Shunde.

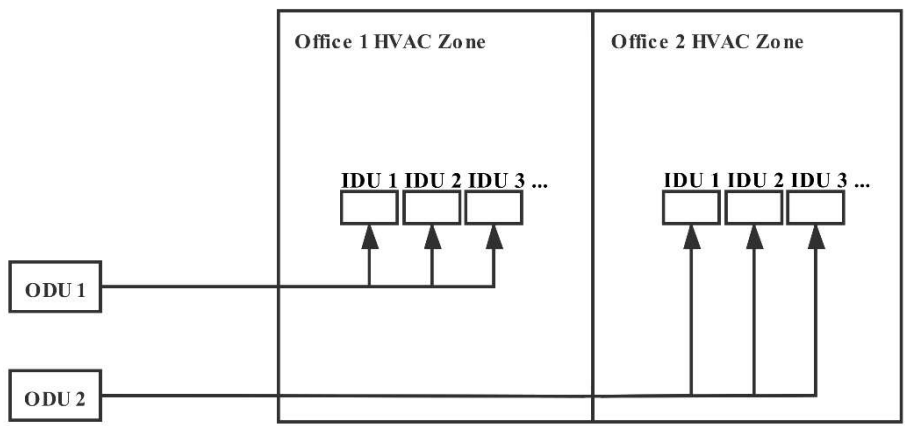

**Fig. 15.** The system layout of the application

The general layout is illustrated in Figure 15, the sector has two HVAC zones, each zone has a VRF system, each VRF system has one indoor unit and three outdoor units. Figure 16 better illustrates the entities and the relationships of the section, a Brick model is created to contain the information. The model is uploaded to the server as a building ontology reference.

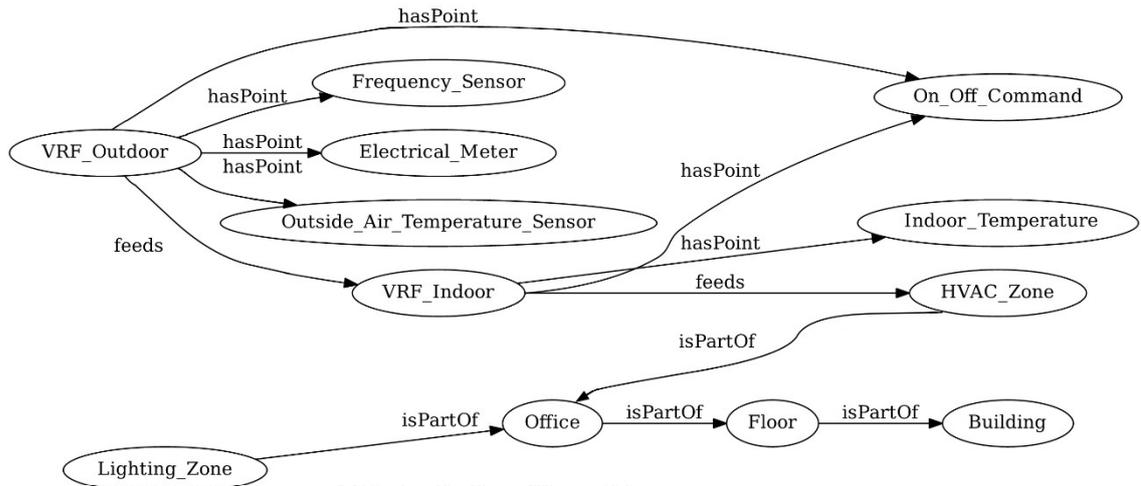

(a) A visualization of the model

```
@prefix b6: <http://example.com/b6#> .
@prefix brick:
<https://brickschema.org/schema/Brick#> .
@prefix b6: <b6:>.
@prefix : <b6:>.
b6:Office_1_HVAC a brick:HVAC_Zone ;
    brick:isPartOf b6:Office_1 .

b6:Office_1_VRF_Indoor_1_On_Off_Command a
brick:On_Off_Command .

b6:Office_1_VRF_Indoor_2_On_Off_Command a
brick:On_Off_Command .

b6:Office_1_VRF_Indoor_3_On_Off_Command a
brick:On_Off_Command .

b6:Office_1_VRF_outdoor_On_Off_Command a
brick:On_Off_Command .

b6:Office_1_lighting a brick:Lighting_Zone ;
    brick:isPartOf b6:Office_1 .

b6:Office_2_HVAC a brick:HVAC_Zone ;
    brick:isPartOf b6:Office_2 .

b6:Office_2_VRF_Indoor_1_On_Off_Command a
brick:On_Off_Command .

b6:Office_2_VRF_Indoor_2_On_Off_Command a
brick:On_Off_Command .

b6:Office_2_VRF_Indoor_3_On_Off_Command a
brick:On_Off_Command .

b6:Office_2_VRF_outdoor_On_Off_Command a
brick:On_Off_Command .
```

```
b6:Office_1_VRF_Indoor_1_T2B a
brick:Temperature_Sensor .

b6:Office_1_VRF_Indoor_2 a brick:VRF_Indoor ;
    brick:feeds brick:Office_1_HVAC ;
    brick:hasPoint b6:Office_1_VRF_Indoor_2_T1,
        b6:Office_1_VRF_Indoor_2_T2,
        b6:Office_1_VRF_Indoor_2_T2B,

brick:Office_1_VRF_Indoor_2_On_Off_Command .

b6:Office_1_VRF_Indoor_2_T1 a
brick:Return_Air_Temperature_Sensor .

b6:Office_1_VRF_Indoor_2_T2 a
brick:Temperature_Sensor .

b6:Office_1_VRF_Indoor_2_T2B a
brick:Temperature_Sensor .

b6:Office_1_VRF_Indoor_3 a brick:VRF_Indoor ;
    brick:feeds brick:Office_1_HVAC ;
    brick:hasPoint b6:Office_1_VRF_Indoor_3_T1,
        b6:Office_1_VRF_Indoor_3_T2,
        b6:Office_1_VRF_Indoor_3_T2B,

brick:Office_1_VRF_Indoor_3_On_Off_Command .

b6:Office_1_VRF_Indoor_3_T1 a
brick:Return_Air_Temperature_Sensor .

b6:Office_1_VRF_Indoor_3_T2 a
brick:Temperature_Sensor .

b6:Office_1_VRF_Indoor_3_T2B a
brick:Temperature_Sensor .
```

(b) Parts of the model in codes

**Fig. 16.** The Brick model of the application

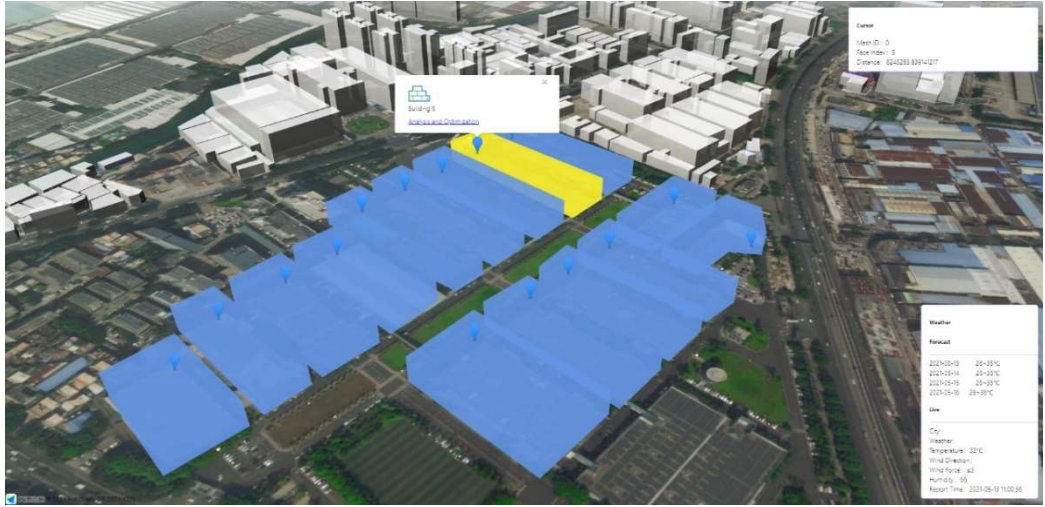

**Fig. 17.** The interaction front-end

A front-end in Figure 18 is built upon the server for ensuring user-friendly. From the interface, the user can select and check relative information. A real-time weather data and weather forecast are attached to the right. When selected, the info window will present a link to the analysis page. The analysis page can retrieve and analyze data from the hosting server. The query that requests for the entity can query the timeseries database simultaneously, which returns data within the query criteria. An instant analysis can be established with PyCaret (Ali 2020), which is an open-source, low code machine learning library in Python. The Brick model provides a well-defined information meaning table so multimedia metadata can be easily combined and processed by humans and applications. Consequently, building management with Brick becomes handy.

The following images are used as a demonstration of building energy analysis. When querying for outside temperature and energy meter records, the server will return data of both VRF systems in Office 1 and Office 2, and data analysis for both rooms can be done at the same time. The results of energy baseline analysis for Office 1 and Office 2 are in Figure 18. Although the overall trend of the forecast results is greatly affected by February and the accuracy needs to be improved, the application in this block shows the availability of the Brick VRF module and the data server.

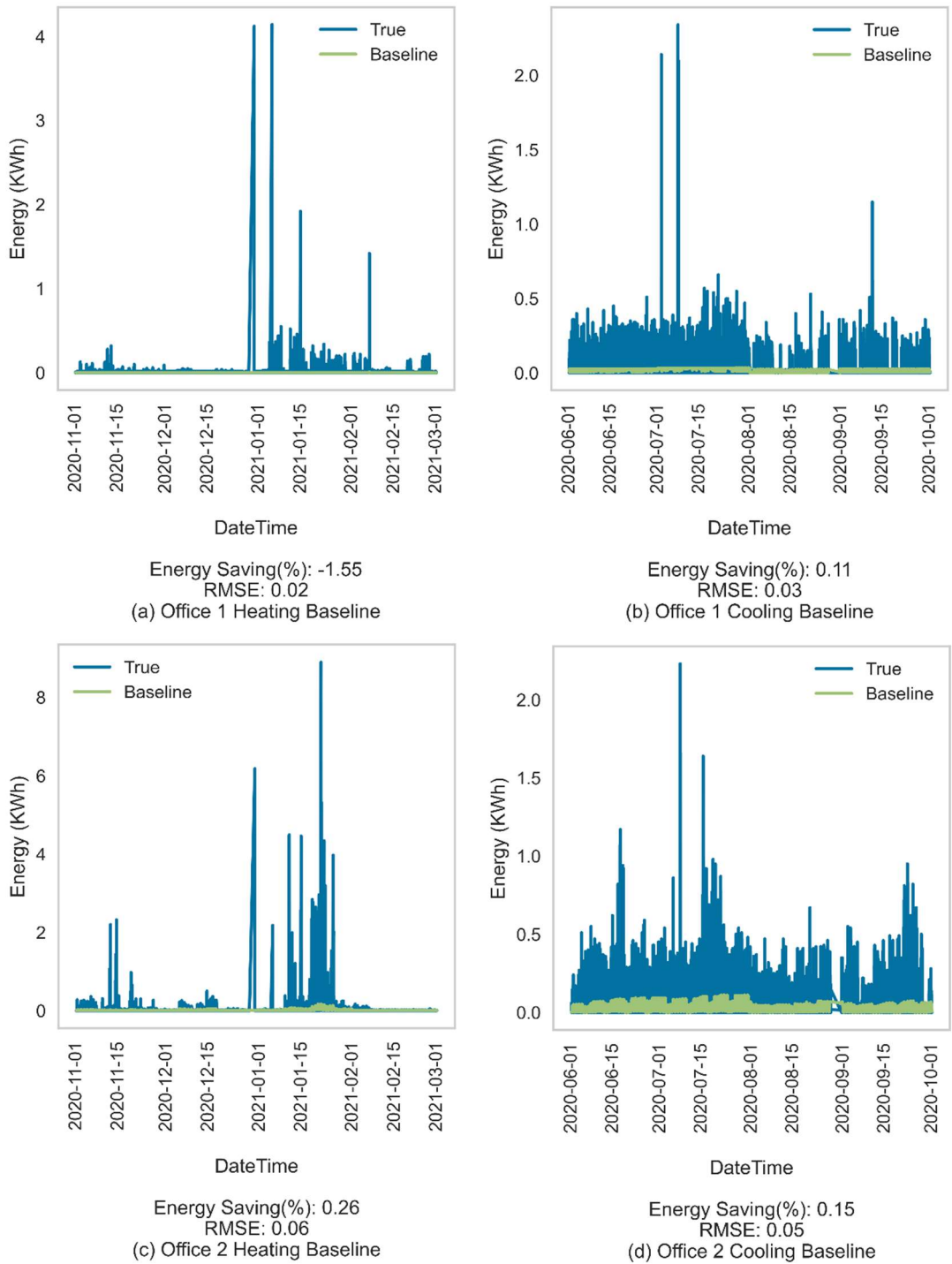

**Fig. 18.** The energy baselines in the application case

## 5. Discussions

The Brick VRF module has been evaluated in the validation and application. After setting the data points of the module and the possible connections with other devices, the Brick VRF module can cooperate with other existing modules due to the scalability of Brick. VRF devices can build the location relationship with the spaces, the supply relationship with the refrigeration area, and the point relationship with the sensor. The multidimensional links in Brick represent the data sources. Instead of storing values of data points, the schema describes connections of all sorts of building databases.

The application case is to evaluate the server's ability to handle various data linked by the model in Figure 13. The results of the query and analysis indicated that data management from a hosting server backed by Brick became reasonable to machines and humans. Besides storing data and hosting queries, because of the information connections being guided by the Brick model, the application can run simultaneous data analysis for different spaces.

However, because the data used in the application has a unified standard, and the building operation data are still stored in a single database, it does not effectively verify Brick's role link in the massive data lake. This problem needs further research in the future.

## 6. Conclusions

In this paper, the Brick VRF module is developed and applied to solve the problem that there is no Hydrofluorocarbon defined in Brick and how to classify VRF equipment in Brick.

This paper presents the VRF module to extend Brick Schema and verifies the Brick VRF module. Then, the model and the service framework are developed and applied for a building sector in China. The application case demonstrates the advantages of Brick in managing multi-building datasets. The VRF module of this paper provides a solution for the expression of the VRF system in the building semantic web.

However, it should be mentioned that the development of the service framework in this paper only uses brick for storage and query, and its advantages in multi-building massive data management need to be further explored, which can be used as future research work.

**Acknowledgments**

This study is supported by HVAC & Building Technologies Division of Midea Group.